\documentclass[aps,showpacs,preprint,superscriptaddress,nofootinbib]{revtex4}
\usepackage{graphicx}
\usepackage{subfigure}
\usepackage{float}
\usepackage{amsmath}
\usepackage{amsfonts}
\usepackage{color}
\usepackage{bm}
\usepackage{bbm}
\usepackage{txfonts}
\usepackage{array}
\usepackage{amssymb}

\begin{document}

	\title{Vortex states and entanglement properties in multiphoton pair production}

	\author{Hong-Hao Fan}
	\affiliation{Key Laboratory of Beam Technology of the Ministry of Education, and School of Physics and Astronomy, Beijing Normal University, Beijing 100875, China}

	\author{Lie-Juan Li}
	\affiliation{School of Mathematics and Physics, Lanzhou Jiaotong University, Lanzhou 730070, China}
	
	\author{Zhi-Hang Yao}
	\affiliation{Key Laboratory of Beam Technology of the Ministry of Education, and School of Physics and Astronomy, Beijing Normal University, Beijing 100875, China}

	\author{Orkash Amat}
	\affiliation{School of Astronomy and Space Science, Nanjing University, Nanjing 210023, China}
		
	\author{Suo Tang}\thanks{tangsuo@ouc.edu.cn}
	\affiliation{College of Physics and Optoelectronic Engineering, Ocean University of China, Qingdao, Shandong, 266100, China}
	
	\author{Bai-Song Xie}\thanks{bsxie@bnu.edu.cn}
	\affiliation{Key Laboratory of Beam Technology of the Ministry of Education, and School of Physics and Astronomy, Beijing Normal University, Beijing 100875, China}
	\affiliation{Institute of Radiation Technology, Beijing Academy of Science and Technology, Beijing 100875, China}

\date{\today}
\begin{abstract}
We investigate the multiphoton pair production in circularly polarized field via two level model. There appears obvious discrete ring structures in the momentum distribution of the created particles, in which the ring radius is mainly controlled by the number of the photons absorbed in the creation with the energy conservation and could also be modulated by the spin of the created pair. These multiphoton rings become narrower when both of the pair particles' spin are aligned with the direction of the field rotation, and become broader if both spin are antiparallel to that direction. This spin-modulation can be simply understood with the angular momentum conservation, as less orbital angular momentum from the absorbed photons would be transferred to the created particles if their spins are aligned with the field rotation. The orbital angular momentum of the created particles is manifested as the vortex structure in the phase of the momentum distribution, and valued as the topological charge of this phase vortex. We also study the spin entanglement between the created particles, and reveal that the entanglement becomes stronger with the increase of the particles' transverse momentum, and gets sharp peak in the transition between different multiphoton rings, where the topological charge of the phase vortex is changed.
\end{abstract}
\pacs{12. 20. Ds, 03. 65. Pm, 02. 60. -x}
\maketitle

\section{Introduction}

Nonlinear quantum electrodynamics (QED) in intense laser pulses has become a hot topic in recent years because of advancements in ultra-intense laser technology. The most striking and fundamental prediction of QED is that the vacuum becomes unstable in the presence of strong external fields, leading to the creation of electron-positron pairs~\cite{Sauter:1931zz,Heisenberg:1936nmg,Schwinger:1951nm,PhysRevD.44.1825,Fradkin:1991,DiPiazza:2011tq,Xie:2017xoj}.
Some recent studies on pair production including the dependence of the momentum spiral on spin~\cite{Hu:2024nyp}, spin/helicity-resolved particle distributions~\cite{Amat:2024nvg,Aleksandrov:2024cqh,Hu:2024nyp,Chen:2025xib}, entanglement~\cite{Li:2016zyv}, vortex state properties~\cite{Fan:2024nsl}, and spiral  structure~\cite{Li:2017qwd} have revealed rich physical phenomena.

It is of great interest to extend the classical vortices to the case of quantum physics, for example the phase vortices, which plays a significant role in various fields ~\cite{berry,PMDirac:1931,BialynickiBirula1992TheoryOQ,Bialynicki-Birula:2016unl} such as molecular vibrations~\cite{Luski:2021}, vortex electron~\cite{Ivanov:2022jzh,Bliokh:2017}, wave vortices~\cite{Bliokh:2025}, and laser-driven quantum dynamics~\cite{Ngoko:2015,Hebenstreit:2016xhn,Pengel:2017,Majczak:2022xlv,Geng2020VortexSI} and so on. Significant progress has been achieved in the theoretical and experimental study of optical vortex beams carrying quantized orbital angular momentum (OAM) as well as electron vortex beams with large OAM~\cite{Sukhorukov:2005,Bliokh:2012,Jhajj:2016,Hancock:2019,Huang:2022}.
Furthermore, in the field of vacuum pair production, vortex-antivortex structures also appear in the probability amplitude of the Sauter-Schwinger process~\cite{Bechler:20232024}. Similarly, under time-dependent external field, the amplitude of photoelectric ionization has a pronounced vortex structure~\cite{Pengel:2017,Majczak:2022xlv,Cajiao:2020}.

In multiphoton scalar pair production, the vortex structures of the probability amplitude are related to the intrinsic OAM of the created particles~\cite{Fan:2024nsl}.
However, in electron-positron pair production with non-zero spin, the angular momentum consists of both spin and orbital contributions, unlike in the multiphoton bosonic case~\cite{Bliokh:2022l}. The vortex structure of the probability amplitude is impacted by spin, which changes the one-to-one correspondence between absorbed photons and topological charges.

Recently, the entanglement properties of electron-positron pair from vacuum has attracted ones great interest~\cite{Balasubramanian:2011wt,Ebadi:2014ufa,Florio:2021xvj,Tang:2025wgm}. In Ref.~\cite{Majczak:2025axk}, the production of helicity-entangled states was investigated, revealing that an external electric field can serve as a switch between different entangled configurations. The feasibility of producing maximally entangled helicity states and employing a short electric pulse as a rapid switching mechanism was also discussed.
Spin not only shapes the phase-vortex structure of the produced particles but also governs their entanglement properties. Although particle pairs created in strong external fields exhibit significant quantum correlations~\cite{Tang:2025wgm}, the emergence of vortex structures introduces additional degrees of freedom associated with OAM. This naturally raises the question of whether phase vortices and entanglement are intrinsically connected.

Motivated by the factors discussed above, it is crucial to explore how spin impact vortex structures in strong external fields and to gain insight into the spin entanglement of the created particles. In this work, we derive the momentum-space particle distributions from the Dirac equation and subsequently reduce the system to a two-level model. Within this framework, the probability amplitude for pair production is expressed as a complex number whose phase factor encodes different spin configurations of the electron and positron. Based on this two-level description, we examine how various spin configurations affect the momentum distributions of the created pair and the associated phase-vortex structures. We further analyze how the corresponding topological charges evolve with the frequency of the external field. In addition, we phenomenologically investigate the entanglement characteristics of the created particles.

Our numerical results show that when the effective magnetic field is introduced, it can be understood why the probability of pair production is largest along the direction of the photon's helicity.
Moreover, different spin configurations lead to distinct topological charges, which are associated with OAM and increase with momentum. For the brightest multiphoton ring dominating pair production, the variation of topological charges with field frequency can be interpreted in terms of energy and angular momentum conservation. In the low-momentum region, the tunneling assisted effect becomes particularly significant. Finally, we find that the entanglement entropy of the produced particles increases with the number of absorbed photons, and becomes especially pronounced in the transition region between the four-photon and five-photon channels.

This paper is organized as follows.
In Sec.~\ref{Sec:II}, the external field and particle distributions are introduced, as well as the description of the vortex and entanglement are reviewed.
In Sec. \ref{Sec:III}, the impact of spin on vortex and the impact of frequency on topological charges are presented.
In Sec.~\ref{Sec.IV}, the entanglement characteristics of the created particles are investigated.
Finally, a brief summary is given in Sec.~\ref{summary}.

\section{Theoretical method}\label{Sec:II}

Pair production through tunneling or multi-photon absorption is typically studied in many works~\cite{Li:2016zyv} where a laser pulse collides with a target or a counter-propagating pulse. In these setting, the electromagnetic invariants $F_{\mu\nu} F^{\mu\nu}$ and $ F_{\mu\nu}\tilde{F}^{\mu\nu}$ are nonzero, enabling pair production~\cite{Ringwald:2001ib}. Usually two counter-propagating laser pulse via head-on colliding can form a standing wave, which is useful to study the electron-positron pair production. Since that the pair production regime is of order of electron Compton scale, which is smaller very much than the wavelength of applied external field, therefore, the spatial dependence of field can be ignored. Now it is convenient to choose the following idealized model of circularly polarized time-dependent field
\begin{align}\label{Eq:1}
	\boldsymbol{E}\left(t\right) = \frac{E}{\sqrt{2}}\exp\left(-\frac{\phi^2}{2\sigma^2}\right) \left[\cos (\phi) \boldsymbol{e_x}  + \delta \sin(\phi) \boldsymbol{e_y}  \right],
\end{align}
where $E $ gives the external field amplitude in units of the critical field $E_\text{cr} = m^2/e \approx 1.3 \times 10^{16} $ V/cm,  $\phi = \omega t$ is the time-dependent carrier phase with frequency $\omega$, and $\sigma$ scales the pulse duration. 
The polarized vectors $ \boldsymbol{e_x} $ and $ \boldsymbol{e_y} $ align with the $ x$- and $ y $-directions, respectively.  The circular polarization corresponds to $ \lvert \delta \rvert = 1 $, with the sign of $ \delta $ determining helicity: $ \delta = 1 $ for right-handed and $\delta = -1 $ for left-handed circular polarization.

\subsection{The description of vortex states}

In our previous work, the electron-positron pair production in momentum-space distributions is derived in Ref.~\cite{Amat:2024nvg}. Therefore, we will not perform a detailed derivation, but will provide a brief review.
Neglecting backreaction, the electron-positron pair production can be obtained in the background electric field $ \boldsymbol{E}\left(t\right) $ given by Eq.~\eqref{Eq:1}, with the associated gauge potential $ \boldsymbol{A}\left(t\right) $, where $ \boldsymbol{E}\left(t\right) = -\dot{\boldsymbol{A}}\left(t\right) $.
A natural formalism for this analysis is provided by Bogoliubov transformations~\cite{Kluger:1998bm}.
The time-dependent creation and annihilation operators for electron spin $ \boldsymbol{s} $ and positron spin $ \boldsymbol{s^\prime} $, denoted as $ \tilde{a}_{\boldsymbol{p,ss^\prime}}(t) $ and $ \tilde{b}_{-\boldsymbol{p,ss^\prime}}^\dagger(t) $, are related to the initial operators $ a_{\boldsymbol{p,ss^\prime}} $ and $ b_{-\boldsymbol{p,ss^\prime}}^\dagger $ via Bogoliubov transformations. 
The transformation coefficients $\alpha_{\boldsymbol{p,ss^\prime}}\left(t\right)$ and $\beta_{\boldsymbol{p,ss^\prime}}\left(t\right)$ characterize this relation as~\cite{Dunne:2022zlx}

\begin{align}\label{Eq:2}
	\begin{bmatrix}
		\tilde a_{\boldsymbol{p,ss^\prime}}\left(t\right) \\
		\tilde b_{-\boldsymbol{p,ss^\prime}}^{\dagger}\left(t\right)
	\end{bmatrix}
	=
	\begin{bmatrix}
		\alpha_{\boldsymbol{p,ss^\prime}}\left(t\right) &  -\beta_{\boldsymbol{p,ss^\prime}}^{*}\left(t\right)\\
		\beta_{\boldsymbol{p,ss^\prime}}\left(t\right) &  \alpha_{\boldsymbol{p,ss^\prime}}^{*}\left(t\right)
	\end{bmatrix}
	\begin{bmatrix}
		a_{\boldsymbol{p,ss^\prime}} \\
		b_{-\boldsymbol{p,ss^\prime}}^{\dagger}
	\end{bmatrix} \ ,
\end{align}
where $\left| \beta_{\boldsymbol{p,ss^\prime}}\left(t\right)  \right|^2$ is the momentum distributions of the created particles with canonical momentum $\boldsymbol{p}$.
The fermi statistics require  $\displaystyle \sum_{ss'} \left[ |\alpha_{p,ss'}(t)|^2 + |\beta_{p,ss'}(t)|^2 \right] = 1$.  The time-dependent Bogoliubov
coefficients satisfy the following relations:
\begin{align}\label{Eq:3}
	\frac{d\alpha_{\boldsymbol{p,ss^\prime}}\left(t\right)}{dt} = \Omega_{\boldsymbol{p,ss^\prime}}\left(t\right) \beta_{\boldsymbol{p,ss^\prime}}\left(t\right) e^{2i\int^t d \tau \omega_{\boldsymbol{p}}(\tau)}, \notag \\
	\frac{d\beta_{\boldsymbol{p,ss^\prime}}\left(t\right)}{dt} = -\Omega_{\boldsymbol{p,ss^\prime}}^{*}\left(t\right) \alpha_{\boldsymbol{p,ss^\prime}}\left(t\right) e^{-2i\int^t d \tau\omega_{\boldsymbol{p}}(\tau)},	
\end{align}
where  $\Omega_{\boldsymbol{p,ss^\prime}}\left(t\right) =  {u_{\boldsymbol{p,s}}^\dagger(t) \dot{H}_{\boldsymbol{p}}(t) v_{\boldsymbol{p,s^\prime}}(t)}/{[2 \omega_{\boldsymbol{p}}(t)]}$, $u_{\boldsymbol{p,ss^\prime}}$ and $v_{\boldsymbol{p,ss^\prime}}(t)$ represent the instantaneous eigenstates of the time-dependent Dirac Hamiltonian $H(t) = \left[\boldsymbol{\alpha}\cdot \left(\boldsymbol{p} - e\boldsymbol{A}(t)\right) + \beta m\right]$ for electron and positron, respectively, and $\omega_{\boldsymbol{p}}\left(t\right)=\sqrt{m^2 +\boldsymbol q^2\left(t\right)} = \sqrt{m^2 + \left[\boldsymbol{p} - e\boldsymbol{A}\left(t\right)\right]^2}$ denotes the particle's energy
with kinetic momentum $\boldsymbol {q}\left(t\right)$ given by  $\boldsymbol{q}\left(t\right) = \boldsymbol{p} - e\boldsymbol{A}\left(t\right)$. The time evolution of the system can be directly expressed  in terms of the Bogoliubov coefficients, whose dynamics are described by a two-level system. By setting $c^{\alpha}_{{\boldsymbol{p,ss^\prime}}}(t)=e^{-i\int^t d \tau \omega_{\boldsymbol{p}}(\tau)}\alpha_{\boldsymbol{p,ss^\prime}}\left(t\right)$, $c^{\beta}_{{\boldsymbol{p,ss^\prime}}}(t) = e^{i\int^t d \tau \omega_{\boldsymbol{p}}(\tau)}\beta_{\boldsymbol{p,ss^\prime}}\left(t\right) $, the evolution equation takes the form
\begin{align}
	i\frac{d}{d t}
	\begin{bmatrix}
		c^{\alpha}_{{\boldsymbol{p,ss^\prime}}}\left(t\right) \\
		c^{\beta}_{{\boldsymbol{p,ss^\prime}}}\left(t\right)
	\end{bmatrix} & =
	\begin{bmatrix}
		\omega_{\boldsymbol{p}}\left(t\right) & i\Omega_{\boldsymbol{p,ss^\prime}}\left(t\right) \\
		-i \Omega_{\boldsymbol{p,ss^\prime}}^{*}\left(t\right) & -\omega_{\boldsymbol{p}}\left(t\right)
	\end{bmatrix}
	\begin{bmatrix}
		c^{\alpha}_{{\boldsymbol{p,ss^\prime}}} \\
		c^{\beta}_{{\boldsymbol{p,ss^\prime}}}
	\end{bmatrix}\label{eq:twolevelsys}.
\end{align}
It is emphasized
that the spin refers to the projection of the electron-positron pair's spin along the $z$-direction,
denoted as $S_z = s_z + s_z^\prime$ for electron (positron) spin projections $s_z$ ($s_z^\prime$),
rather than the two-particle singlet or triplet states. The spin-dependent particle distributions  in momentum space are expressed as
\begin{align}\label{Eq:5}
	f_{\boldsymbol{ss^\prime}}({\boldsymbol{p}}) = 2\lvert c^{\beta}_{{\boldsymbol{p,ss^\prime}}} \rvert ^2.
\end{align}
The particle distributions of the total created particles can be obtained by summing the spins:
\begin{align}\label{Eq:6}
	f = \sum_{{\boldsymbol{ss^\prime}}} f_{\boldsymbol{ss^\prime}}.
\end{align}

It is well know that the vortex lines are closed loops or continuous curves in three-dimensional momentum space~\cite{Majczak:2022xlv,Hebenstreit:2016xhn,Bliokh:2011fi,Bliokh:2012az}. The nodes form surfaces with vanishing probability, where the phase of the probability amplitude jumps $\pm \pi$~\cite{Geng2020VortexSI}.
Note that in the two-dimensional plane, vortex lines are usually visualized as individual points, while nodal surfaces behave as curves with zero probability~\cite{Majczak:2022xlv,Cajiao:2020}.
The complex probability amplitude $ c^{\beta}_{{\boldsymbol{p,ss^\prime}}}$
in the external electric field contains essential physical information. We know that the spin-dependent Berry connection can be defined as~\cite{Hebenstreit:2016xhn,Geng2020VortexSI,Bechler:20232024}
\begin{align}\label{Eq:7}
	\mathcal{A}_{\boldsymbol{ss^\prime}} = \frac{\text{Re} \left[ \left(c^{\beta}_{{\boldsymbol{p,ss^\prime}}}\right)^* \left(-i \nabla_{\boldsymbol{p}} \right)  c^{\beta}_{{\boldsymbol{p,ss^\prime}}} \right]}{\lvert  c^{\beta}_{{\boldsymbol{p,ss^\prime}}} \rvert^2} = \nabla_{\boldsymbol{p}} \left( \arg \left[  c^{\beta}_{{\boldsymbol{p,ss^\prime}}} \right] \right),
\end{align}
and it is connected to the quantization condition
\begin{align}\label{Eq:9}
	\frac{1}{2\pi }\oint_\mathcal{C} \mathcal{A}_{\boldsymbol{ss^\prime}} \cdot d\boldsymbol{p} = l_{\boldsymbol{ss^\prime}},
\end{align}
where $ l_{\boldsymbol{ss^\prime}} = 0,\pm 1, \pm2, ..., $ are the spin-dependent winding number or topological charges. This result highlights the relation between the phase structure and the topological properties of the system for different pair spin.

\subsection{Pair entanglement}

In this work, we do not directly extract the entangled states from the full wave function about electron/positron. Instead, we adopt a phenomenological framework based on the spin-resolved pair production probabilities. By utilizing the dependence of the production probability on different spin configurations, the final-state "wave function" of the created particles can be phenomenologically expanded in the spin-projection basis~\cite{Kohlfurst:2018kxg,Dai:2019nzv,Majczak:2025axk}:
\begin{align}\label{Eq:10} |\psi \rangle \propto \left(c^{\beta}_{\uparrow\uparrow}| \uparrow\uparrow\rangle + c^{\beta}_{\downarrow\downarrow}|\downarrow\downarrow\rangle +c^{\beta}_{\uparrow\downarrow}|\uparrow\downarrow\rangle + c^{\beta}_{\downarrow\uparrow}|\downarrow\uparrow\rangle\right).
\end{align}
The squared magnitudes of these coefficients correspond to the pair production probabilities in the respective spin states.

The entanglement of the created particles can be quantified using the entanglement entropy \cite{Bennett:1996gf}
\begin{align}\label{Entropy}
	S(\mathcal{C}) = h\!\left[\frac{1}{2}\big(1 + \sqrt{1 - \mathcal{C}^2}\big)\right],
\end{align}
where $	h(x) = -x\log_2 x - (1 - x)\log_2(1 - x)$, and $\mathcal{C}$ is the concurrence, see Eq.~\eqref{Concurrence}.
To compute $\mathcal{C}$, we first construct the density matrix based on Eq.~\eqref{Eq:10}:
\begin{align}
	\rho = \frac{1}{N(\boldsymbol{p})}
	\begin{pmatrix}
		c_{uu}c_{uu}^* & c_{uu}c_{ud}^* & c_{uu}c_{du}^* & c_{uu}c_{dd}^* \\
		c_{ud}c_{uu}^* & c_{ud}c_{ud}^* & c_{ud}c_{du}^* & c_{ud}c_{dd}^* \\
		c_{du}c_{uu}^* & c_{du}c_{ud}^* & c_{du}c_{du}^* & c_{du}c_{dd}^* \\
		c_{dd}c_{uu}^* & c_{dd}c_{ud}^* & c_{dd}c_{du}^* & c_{dd}c_{dd}^*
	\end{pmatrix},
\end{align}
where $N(\boldsymbol{p})$ is a normalization factor ensuring $\text{Tr}\,\rho = 1$.
The concurrence, introduced in~\cite{Hill:1997pfa,Wootters:1997id}, is then defined as
\begin{equation}\label{Concurrence}
	\mathcal{C} = \max \Big(0, \sqrt{e_1} - \sqrt{e_2} - \sqrt{e_3} - \sqrt{e_4} \Big),
\end{equation}
where $e_j$ $(j = 1,2,3,4)$ are the eigenvalues, in descending order, of the matrix
\begin{equation}
	R = \rho (\sigma_2 \otimes \sigma_2)\, \rho^* (\sigma_2 \otimes \sigma_2),
\end{equation}
with $\rho^*$ being the complex conjugate of $\rho$ and $\sigma_2$ is the second Pauli matrix.
By inserting Eq.~\eqref{Concurrence} into Eq.~\eqref{Entropy}, the entanglement entropy can be calculated. The values of $S(\mathcal{C})$ and concurrence $\mathcal{C}$ range from $0$ to $1$. A value of $S(\mathcal{C}) = 1$ with $\mathcal{C} = 1$ corresponds to a maximally entangled state, while $S(\mathcal{C}) = 0$ with $\mathcal{C} = 0$ denotes a completely separable state, which can be represented as a product of two independent subsystems.

\section{Vortex properties}\label{Sec:III}

In pair production, the Keldysh adiabatic parameter  $\gamma = {m\omega}/({eE}) $ characterizes two different mechanisms as the non-perturbative tunneling-Schwinger effect when $\gamma \ll 1$ and perturbative multiphoton process when $\gamma \gg 1$~\cite{Keldysh:1965ojf}. When we give a fixed frequency $\omega$, the created particles absorb a minimum number of photons
$n_\text{min}$.
In addition, absorption of $n_\text{min} + k$ photons ($k = 0, 1, 2, \dots$) can occur for the created particles with large momentum~\cite{Kohlfurst:2018kxg},
a phenomenon analogous to above-threshold ionization in atomic physics~\cite{Popov:1972,Muller:2009zzf,Popov:2005rp}. In the following numerical calculation, the parameters $\sigma = 20$ and $E = 0.05E_\text{cr}$ are fixed.

\subsection{Impact of pair spin on the vortex}\label{Sec:IIIA}

In multiphoton pair production with $\omega = 0.55m$, the Keldysh adiabatic parameter is $\gamma = 11 \gg 1$, which indicate that the pair are produced by absorbing at least four photons from external field~\cite{Aleksandrov:2024sub,Kudlis:2025stb,Blinne:2013via}. When each photon is absorbed, it transfers its energy to pair, as well as transfers its spin to pair angular momentum.

\begin{figure}[htbp]
	\centering
	\includegraphics[scale=0.5]{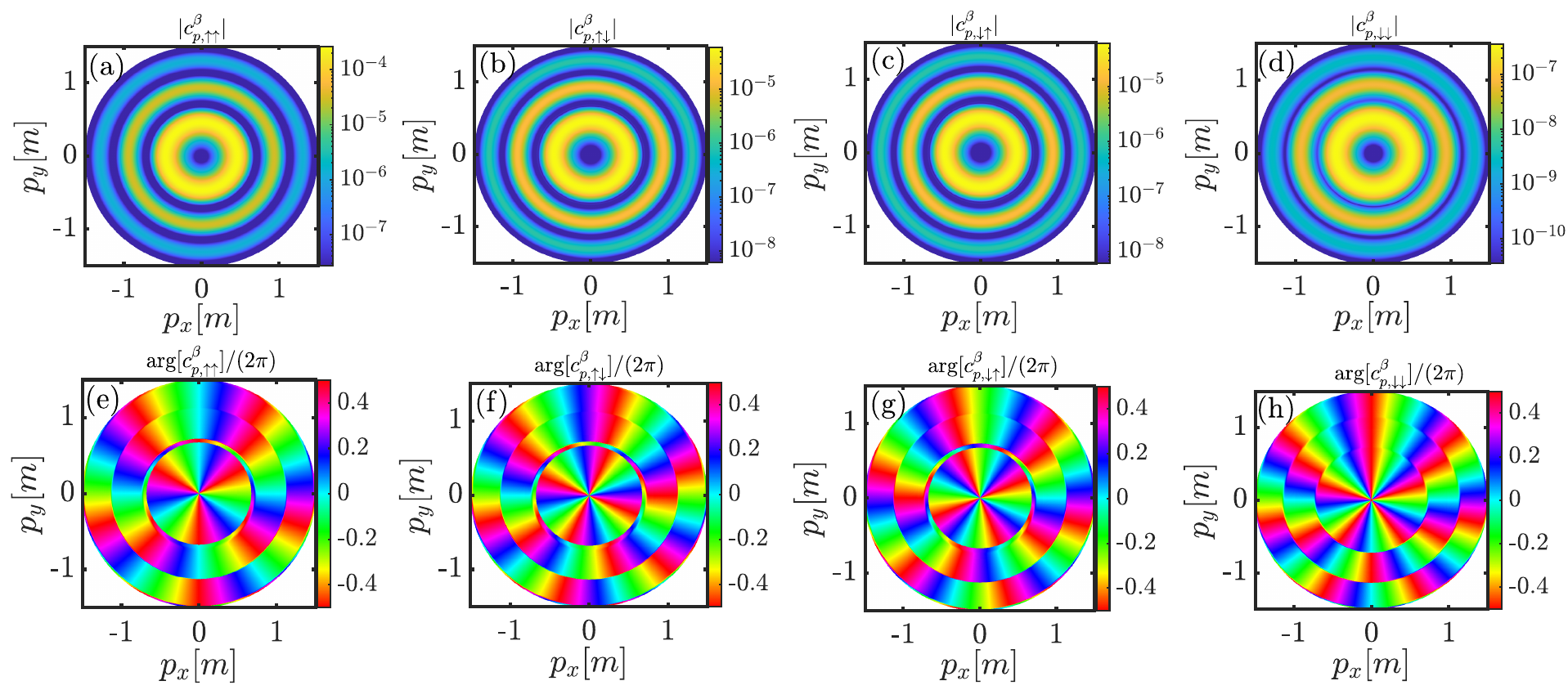}
	\caption{Spin-dependent particle distributions (upper panel) and phase distributions arg$\left[  c^{\beta}_{{\boldsymbol{p,ss^\prime}}} \right]/2\pi$ (lower panel). The parameters are $\omega = 0.55 m$, and  $\delta = 1$.}
	\label{Fig:1}
\end{figure}

In Fig.~\ref{Fig:1}, we give the the particle distributions and the phase distribution in momentum space. From left to right, the spin configurations of the created pair are $S_z = 1$, 0, 0, and $-1$, respectively. Figs.~\ref{Fig:1}(a)-\ref{Fig:1}(d) present the particle distributions of the created particles, and the corresponding phase distributions for different spin configurations are displayed in Figs.~\ref{Fig:1}(e)-\ref{Fig:1}(h). In Figs.~\ref{Fig:1}(a)-\ref{Fig:1}(d), the first (innermost) multiphoton ring corresponds to $k = 0$, the second to $k = 1$, and higher-order rings correspond to increasing values of $k$. Since the angular momentum transfer for $k > 0$ exhibits behavior similar to that for $k = 0$, we focus on the $k = 0$ case in the following discussion.

The results demonstrate that the particle distributions exhibit clear spin dependence: the production probability is highest when the spin projection aligns with the helicity of the external field. This behavior can be understood through the concept of an effective magnetic field.
For a circularly polarized field, taking the electron for example, the effective magnetic field is defined as by~\cite{Chen:2025xib,Metodiev:2015gda}
$
\boldsymbol{B}_\text{eff} = \frac{1}{\omega_{\boldsymbol{p}} + m} \boldsymbol{E} \times \boldsymbol{q}.
$
The momentum of the created particles in the slowly varying envelope approximation can be obtained as $\boldsymbol{q} \approx -\frac{eE}{\sqrt{2}}\exp\left(-\frac{\phi^2}{2\sigma^2}\right) \left[ \sin \left(\phi\right) \boldsymbol{e_x} - \delta \cos\left(\phi\right) \boldsymbol{e_y} \right]$ via $d\boldsymbol{q}/dt = -e\boldsymbol{E}(t)$~\cite{Heinzl:2020ynb}, whose sign is determined by the parameter $\delta$. Both momentum and the external field are located at the ($p_x, p_y$) plane, and thus the effective magnetic field is non-zero only in the $z$ component whose sign is determined by $\delta$, i.e., $\boldsymbol{B}_\text{eff} \propto 0.5{\delta E^2}\exp\left(-\frac{\phi^2}{\sigma^2} \right)\boldsymbol{e_z}$. At this point, the electron spin has a higher probability of aligning along the direction of the magnetic field.
The electron spin therefore tends to align with the direction of this effective magnetic field, resulting in enhanced pair-production probability when the field helicity and spin projection are aligned.

As $\delta$ approaches zero, the effective magnetic field weakens, and the difference in production rates between different spin configurations nearly vanishes, particularly in the case of linearly polarized conditions.

In Fig.~\ref{Fig:1}(a), the spin configuration with $S_z = 1$ exhibits the highest probability, corresponding to both particle spins aligned parallel to the effective magnetic field. In Figs.~\ref{Fig:1}(b) and~\ref{Fig:1}(c), corresponding to $S_z = 0$, one particle's spin is antiparallel to the effective magnetic field, resulting in a lower production probability compared to the $S_z = 1$ case. These configurations represent a spin-exchange scenario: either the electron spin aligns with the effective magnetic field while the positron spin opposes it, or vice versa. The results indicate that the production probability is invariant under this spin exchange, except for a change in the phase factor (see below). When $S_z = -1$, both particle spins are antiparallel to the effective magnetic field, yielding the lowest production probability, as shown in Fig.~\ref{Fig:1}(d).

In Figs.~\ref{Fig:1}(e)-\ref{Fig:1}(h), we find that the vortex structure is clockwise. Moreover, the change of the chirality in the external field can also causes the counterclockwise vortex structure and a spin reversion.
In the field model described by Eq.~\eqref{Eq:1}, the photons do not possess helical wavefront structure, so they do not have intrinsic OAM. Utilizing the conservation of angular momentum, we obtain as~\cite{Li:2017qwd}
\begin{align}\label{Eq:14}
n = l_{\boldsymbol{s}\boldsymbol{s}^\prime} + S_z.
\end{align}
According to Eq.~\eqref{Eq:14}, in the cases of $S_z = +1$ and $S_z = -1$, there are three and five topological charges, respectively, see Figs.~\ref{Fig:1}(e) and~\ref{Fig:1}(h).
In the cases of $S_z = 0$, the topological charges equal the number of absorbed photons, $l_{ \uparrow\downarrow} = l_{\downarrow\uparrow} = 4$, as shown in Figs.~\ref{Fig:1}(f) and~\ref{Fig:1}(g).
Moreover, in a fermion system, a minus sign appears in probability amplitude when the two particles are exchanged odd times, i.e., $c^{\beta}_{{\boldsymbol{p},\uparrow\downarrow}} = - c^{\beta}_{{\boldsymbol{p},\downarrow\uparrow}}$. Therefore, in Figs.~\ref{Fig:1}(b) and~\ref{Fig:1}(c), the peak of the particle distributions remain the same, but the phase shifts by $\pm \pi$.

\begin{table}[htbp]
	\caption{Ring radius for $k = 0$ and corresponding topological charges for different combinations of pair spin.}
	\label{TableI}
	\begin{ruledtabular}
		\begin{tabular}{ccccc}
			pair spin & $\uparrow\uparrow$ & $\uparrow\downarrow$ & $\downarrow\uparrow$ & $\downarrow\downarrow$ \\
			\hline
			$p_r$ & $0.46m$ & $0.47m$ & $0.47m$ & $0.48m$ \\
			$l_{\boldsymbol{ss^\prime}}$ & 3 & 4 & 4 & 5 \\
		\end{tabular}
	\end{ruledtabular}
\end{table}

From Fig.~\ref{Fig:1}, one can see that the larger topological charges correspond to the larger ring radius $p_{r}$ in the particle distributions. The results that the multiphoton ring radius differs in different spin projections can also be understood according to Eq.~\eqref{Eq:14}.  For smaller $S_z$, the angular momentum conservation must be satisfied with the larger OAM, which corresponds to lager ring radius. Thus, as shown in Table~\ref{TableI}, when the spin $S_z = 1$, the topological charges are three and the radius is smallest.  For the cases of $S_z = 0$, the topological charges are four, and the radii of the rings are equally medium.  When the spin $S_z = -1$, the topological charges are five, and the radius is largest.

\subsection{The impact of frequency on spin-dependent topological charge}\label{SecIII:B}

\begin{figure}[htbp]
	\centering
	\includegraphics[scale=0.7]{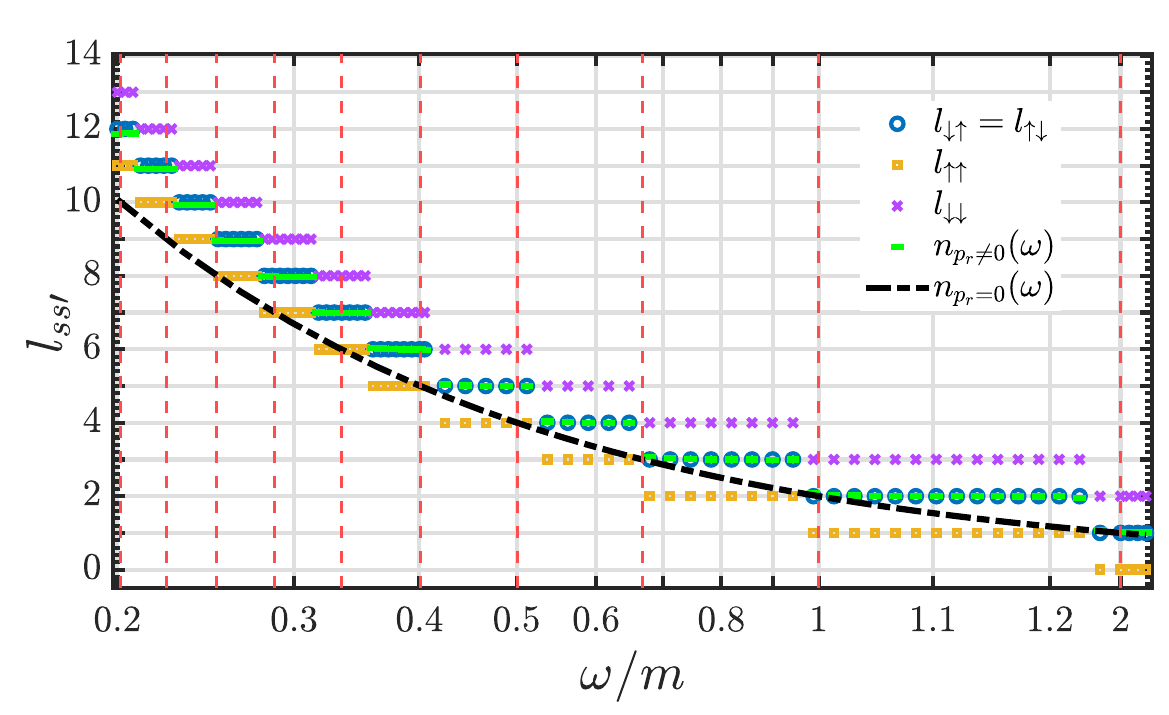}
	\caption{The relationship between the topological charges for the brightest ring and frequency. The blue symbol denoting $S_z = 0$ configurations, as well as the yellow and purple symbols corresponding for $S_z = 1$ and $-1$, respectively. The green symbol represents the number of the photons obtained from Eq.~\eqref{Eq:16} as a function of frequency, where $p_r$ is the radius of the brightest multiphoton ring. The black curve shows the number of the absorbed photons $n_{p_r=0 }(\omega)$ when set the momentum ${p}_{r}=0$. The parameters $E = 0.05E_{\text{cr}}$, and $\delta = 1$ keep constant.}
	\label{Fig:2}
\end{figure}

We now investigate how the topological charges change with the external field frequency. From the above discussion, we know that topological charges are related to the number of absorbed photons. The energy of the created particles is inherited from the photons and can be written as
\begin{equation}\label{Eq:16}
	\mathcal{E}(\boldsymbol{p}) = n\omega/2 = \sqrt{m^2_* + \boldsymbol{p}^2},
\end{equation}
,
where the effective mass is $m_* = m \sqrt{1 + {1}/{(2\gamma)^2}}$. This expression imply that the momentum determine the number of absorbed photons for fixed $\omega$. We select the brightest to investigate the impact of frequency on the topological charges. The topological charges in different spin configurations can be obtained by integrating along the ring using Eq.~\eqref{Eq:9}.
By inserting the momentum radius $p_{r,n}$ of the $n$-photon brightest ring in the particle distribution into Eq.~\eqref{Eq:16},  the corresponding number of absorbed photons $n$ can also be obtained.

\begin{figure}[htbp]
	\centering
	\includegraphics[scale=0.5]{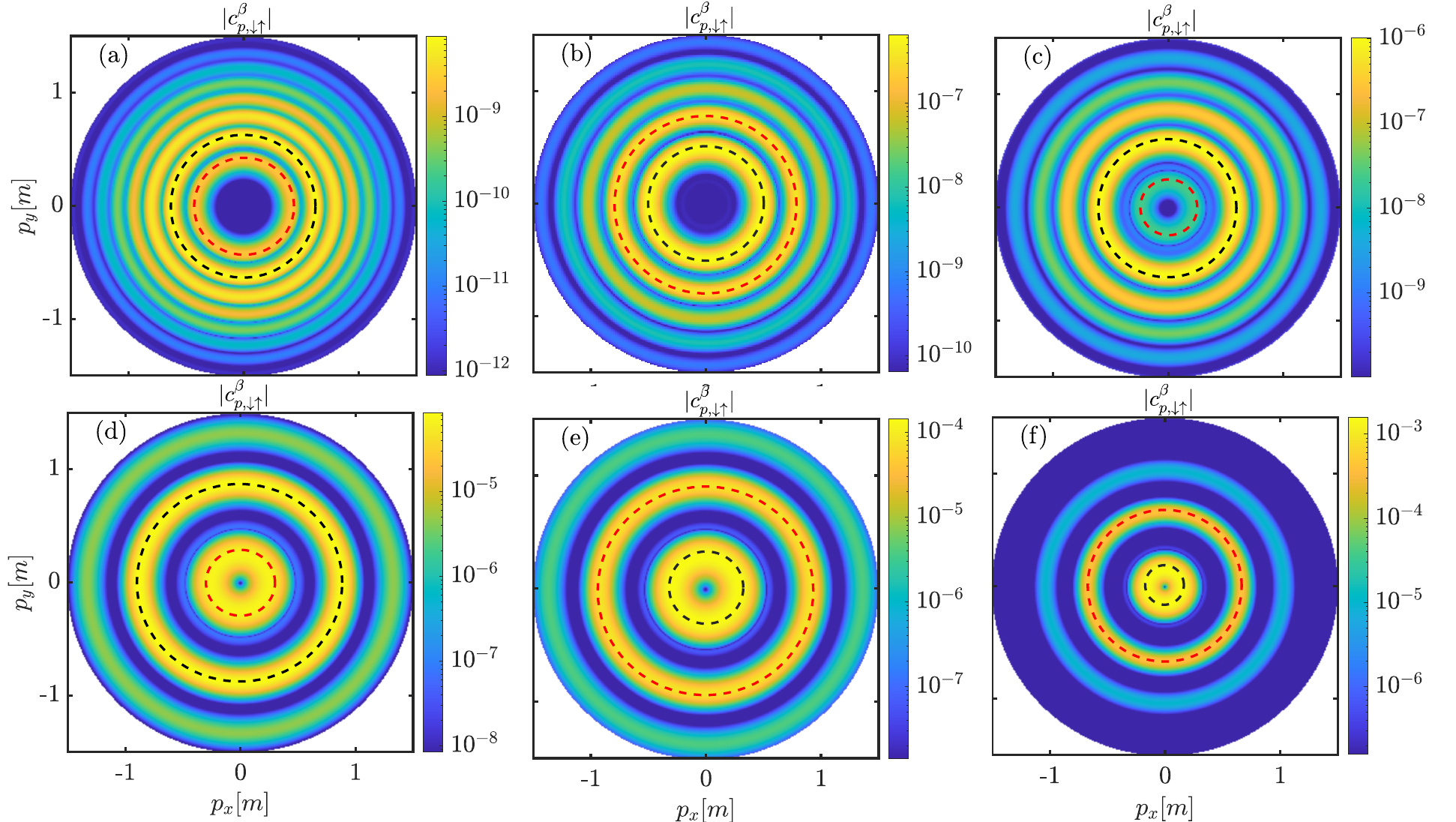}
	\caption{Particle distributions for the case of $S_z = 0$ with the electron spin down and positron spin up. From (a) to (f), the frequencies are  $\omega_n = 0.2m$, $0.32m$, $0.334m$, 0.667$m$, 0.68$m$, $1m$, and the absorbed photons for the brightest rings correspond to 12, 7, 7, 4, 3, 2, respectively. The black and red circles represent the rings with the highest probability and the next highest probability. The parameters $E = 0.05E_{\text{cr}}$, and $\delta = 1$ are fixed.}
	\label{Fig:3}
\end{figure}

In Fig.~\ref{Fig:2}, we show the variation of topological charges with frequency. The red vertical dashed lines, $\omega_n = 2 m_*/n$, correspond to integers $n$ decreasing from 10 to 1 from left to right. It can be seen that the topological charges for different spin configurations exhibit a step-like structure, with the height difference between neighboring steps for the same spin configuration given by $\Delta l_{\boldsymbol{s}\boldsymbol{s}^\prime} = \pm 1$. As the frequency increases, the photon energy increases and the number of absorbed photons decreases, resulting in a reduction of the angular momentum carried by the created particles. At given frequency, particles in the $S_z = -1$ spin configuration (purple symbol) possess the largest OAM compared to other spin configurations, which results in the highest plateau for $l_{\downarrow\downarrow}$.

Interestingly, the step-structure of the topological charges $l_{\boldsymbol{s},\boldsymbol{s}}$ closely relates the variation in the number of absorbed photons $n_{p_r \neq 0}(\omega)$, as shown by the green symbol in Fig.~\ref{Fig:2}. The abrupt changes in $l_{\boldsymbol{s},\boldsymbol{s}^\prime}$ occur at frequencies where $n_{p_r \neq 0}(\omega)$ also exhibits discontinuities, indicating that the OAM is directly linked to the photon's number at a given frequency. When new absobed channels open, the photon number undergoes sudden jumps, and the OAM change accordingly. Such behavior is a direct consequence of angular-momentum conservation.

In Fig.~\ref{Fig:2}, we also observe that the difference between the black curve and the green symbols gradually decreases with increasing external field frequency. This behavior originates from the defference in the structure of particle distributions.
In Fig.~\ref{Fig:3}, we present the particle distributions for external field frequencies $\omega = 0.2m$, $0.32m$, $0.334m$, $0.667m$, $0.68m$, and $1m$. The distributions show that between any two adjacent bright rings there exists a dark ring where the probability nearly vanishes, forming a nodal surface in momentum space. The spacing between adjacent rings is $\Delta p = p_{r,n} - p_{r,(n-1)}$. It can be shown that $\partial \Delta p / \partial \omega > 0$, implying that the rings become sparser in particle distributions when $\omega$ grow.

The above distributions can be classified into two categories.~The first category includes~Figs.~\ref{Fig:3}(a)-~\ref{Fig:3}(d), where the brightest $n$-photon absorption ring is accompanied by a weaker $(n-1)$-photon absorption ring. The second category includes Figs.~\ref{Fig:3}(e) and \ref{Fig:3}(f), where the innermost ring in the distributions correspond to the largest $n$-photon absorption ring. In the first case, the brightest ring occurs in an $n$-photon process, and $n = l_{\downarrow\uparrow} > 2m_*/\omega$, which explains why the black curve lies below the green symbol in Fig.~\ref{Fig:2}. For instance, at low frequency with $\omega = 0.2m$, the brightest ring is the twelve-photon absorption ring (marked by the black circle), followed by a weaker eleven-photon ring (red circle). The topological charges $l_{\downarrow\uparrow}$ obtained by integrating along the black circle is twelve. This value remains constant as long as the integration contour is shrunk without crossing the nodal surface.

When the frequency is increased to $\omega = 0.32m$, the dominant process is identified as seven-photon absorption. In this regime, if the integration contour does not cross any nodal surface, the topological charges remain constant at $l_{\downarrow\uparrow} = 7$.
At this point, the momentum radius is smaller than that of the twelve-photon case, and lead to the discrepancy between $n_{p_r \neq 0}(\omega)$ and $n_{p_r = 0}(\omega)$ decreases as expected in Fig.~\ref{Fig:2}. Similarly, as the frequency increases to $\omega = 0.334m$ and 0.667$m$, the discrepancy between $n_{p_r \neq 0}(\omega)$ and $n_{p_r =0}(\omega)$ continues to decrease.

For the second category,
when the integral is performed over an infinitesimal momentum radius, the topological charges $l_{\downarrow\uparrow}$ remain constant. In this case, the gap between the black and green lines further smaller, see the black curve and the green symbol in Fig.~\ref{Fig:2}.

\subsection{Tunneling assisted effect}

\begin{figure}[htbp]
	\centering
	\includegraphics[scale=0.7]{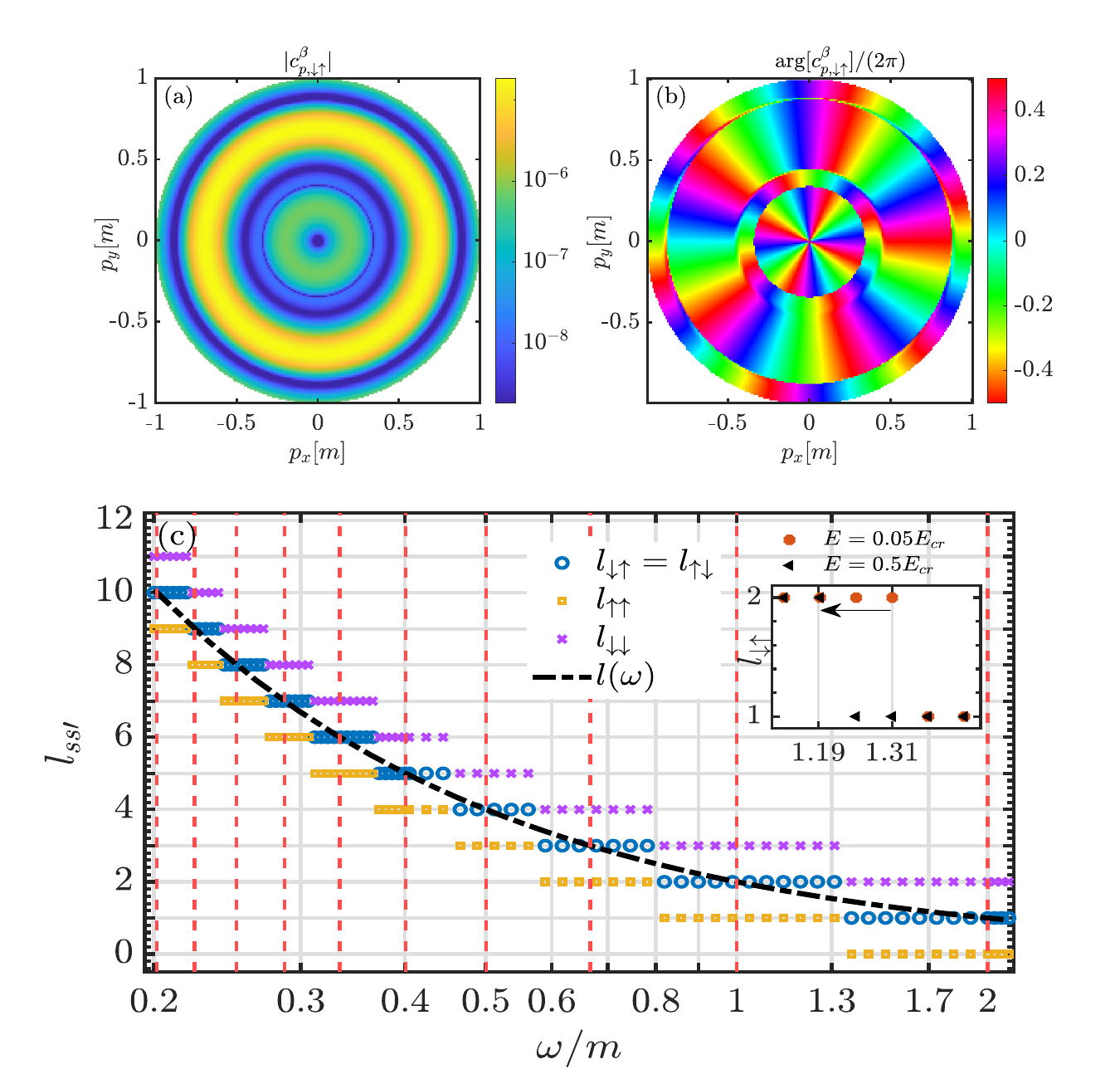}
	\caption{(a) and (b) corresponding to the particle distributions of the created particles and phase distributions for $\omega = 0.488m$, $E = 0.05E_{\text{cr}}$, and $\delta = 1$. (c) same as in Fig.~\ref{Fig:2} but for $p_r = 0.05m$, and the inset gives the topological charges $l_{\downarrow\uparrow}$ variation with frequency for $E = 0.05 E_\text{cr}$ with red symbols and $E = 0.5E_\text{cr}$ with blue symbol.}
	\label{Fig:4}
\end{figure}

It is noteworthy that the absorption mechanism in the firs category exhibits fundamental differences compared to those in the second.
In Figs.~\ref{Fig:4}(a) and~\ref{Fig:4}(b), we display the particle distributions and phase distributions for electron spin down and positron spin up at $\omega = 0.488m$.
The particle distributions reveal a weak ring, and the phase distributions show that this ring structure is accompanied by four topological charges that is a character of a four-photon absorption process.
This structure resembles the patterns observed in Figs.~\ref{Fig:3}(a)-~\ref{Fig:3}(d).

In the sub-threshold regime where $n\omega < 2m_*$, if the $n$ topological charges can be observed, $n$-photon pair production can occur below the energy threshold.
According to the effective mass model, five-photon absorption would be appeared at $\omega \approx 0.488m$.
However, weak tunneling assisted may impact the multiphoton absorption~\cite{Blinne:2016yzv}.
In this case, tunneling assistance becomes essential, that is, if the energy of $n$ photons is insufficient to produce real particles, the external electric field supplies the remaining energy through its work, resulting in pair creation.
When $n\omega > 2m_*$,  the pattern of the particle distributions recover the structure shown  Figs.~\ref{Fig:1}(a)-\ref{Fig:1}(d).

In Fig.~\ref{Fig:4}(c), we select a small momentum with modulus $ p_r = 0.05m $ and use Eq.~\eqref{Eq:9} to determine topological charges at various frequencies in different spin configurations.
The dashed red lines and black dashed curve are same as in Fig.~\ref{Fig:2}.
Due to angular momentum conservation, intersection point of  the black and red lines lie in the step of $l_{\downarrow\uparrow}$ for the $n$-photon absorbed channel. As the frequency increases, the intersection point far away from the left edge of the fixed step in $n$-photon absorbed channel.
Furthermore, stronger electric fields enhance tunneling assisted effect, causing the step structure to shift toward to left.
The inset in Fig.~\ref{Fig:4}(c) depicts the variation of topological charges $ l_{\uparrow\downarrow} $ with frequency, using $ E = 0.5E_\text{cr} $ as an example to highlight the tunneling assisted effect. It can be found that when the field strength increases by a factor of ten, the step shifts to the left as shown the black arrow, that is, the frequency at which the new two-photon channel opens has shifted from $\omega = 1.31m$ to $1.19m$.

\section{Entangled characteristics}\label{Sec.IV}

\begin{figure}[htbp]
	\centering
	\includegraphics[scale=0.71]{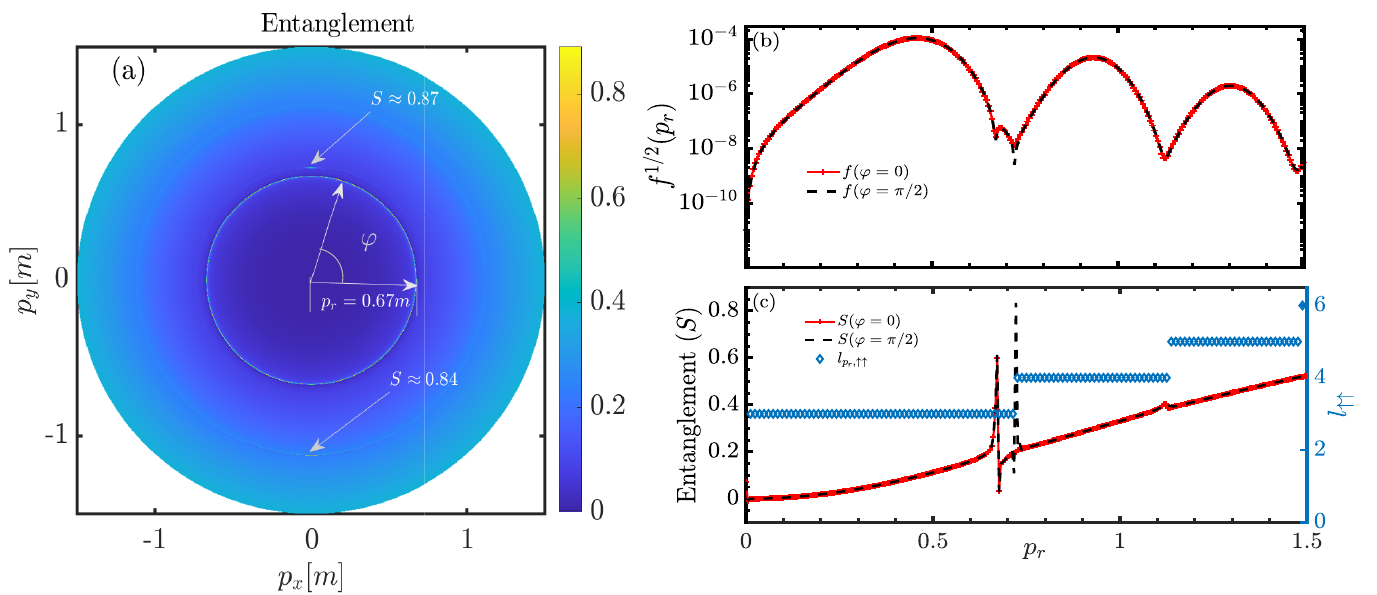}
	\caption{(a) Entanglement entropy in the transverse momentum plane $(p_x, p_y)$.
		(b) Radial distributions of $f^{1/2}(p_r)$ for $\varphi = 0$ and $\varphi = \pi/2$.
		(c) Radial distributions of the entanglement entropy and topological charges at $\varphi = 0$ and $\varphi = \pi/2$.  The parameters $E = 0.05\,E_{\text{cr}}$,  $\omega = 0.55m$, and  $\delta = 1$ are kept constant.}
	\label{Fig:5}
\end{figure}

We now investigate the entanglement properties of the created electron-positron pair.
As shown in Figs.~\ref{Fig:1}(e) and~\ref{Fig:1}(f), the phase difference between the $S_z = +1$ and $S_z = -1$ spin channels exhibits a pronounced dependence on momentum. Phenomenologically, the final-state "wave function" of the produced pair can be expressed as
$|\psi\rangle \propto
\left[c^{\beta}_{\uparrow\downarrow} \bigl( |\!\uparrow\downarrow\rangle - |\!\downarrow\uparrow\rangle \bigr)
+ c^{\beta}_{\uparrow\uparrow} |\!\uparrow\uparrow\rangle
+ c^{\beta}_{\downarrow\downarrow} |\!\downarrow\downarrow\rangle\right]$,
The entanglement entropy of the created particles can then be evaluated by Eq.~\eqref{Entropy}.

Fig.~\ref{Fig:5}(a) shows the momentum-space distribution of the particle entanglement entropy $S$. At $p_r \approx 0.67m$, the entanglement reaches approximately $S \approx 0.8$ and is nearly independent of the azimuthal angle $\varphi$, where $\varphi$ is measured from the $p_x$ axis. A second region of strong entanglement appears at $p_r \approx 0.724m$ and $\varphi = \pi/2$, where $S \approx 0.87$ (see also Fig.~\ref{Fig:5}(c)). Similarly, another peak with $S \approx 0.84$ is observed at $p_r \approx 1.124m$ and $\varphi = 3\pi/2$. These regions of strong entanglement occur in the transitional domains between four- and five-photon absorption, as well as between five- and six-photon absorption. A change in the number of absorbed photons leads to an increase in entanglement entropy.

In Fig.~\ref{Fig:5}(b), we present the total particle yield as a function of the radial momentum $p_r$ for azimuthal angles $\varphi = 0$ and $\varphi = \pi/2$. The radial distributions are nearly identical for these two angles. Each valley in the distributions (except the first) marks the opening of a new multiphoton absorption channel, while the peaks correspond to active multiphoton absorption processes. At the second valley, the yield for $\varphi = \pi/2$ is slightly lower than that for $\varphi = 0$, indicating the presence of a node at this location. These valley positions coincide with the regions of high entanglement entropy and correspond to newly opened multiphoton absorption channels. Within these regions, the increase in the number of absorbed photons indicates that the spin entanglement is affected by the photon absorption process. An increase in the number of absorbed photons leads to an increase in entanglement entropy.

In Fig.~\ref{Fig:5}(c), we present the radial dependence of the topological charges $l_{\uparrow\uparrow}$ (blue step) and the entanglement entropy $S$ for the azimuthal angles $\varphi = 0$ and $\varphi = \pi/2$.
Except near $p_r \approx 0.724m$, the entanglement entropy is nearly identical for $\varphi = 0$ and $\varphi = \pi/2$, with the two curves almost completely overlapping.
The radial profile of $S$ shows that the entanglement generally increases with momentum, apart from a sharp peak followed by a rapid drop at the valley, which is consistent with the presence of a nodal structure. The step structure in the topological charges $l_{\uparrow\uparrow}$ arises from the increasing number of absorbed photons at higher momenta, which enhances the OAM carried by the produced pairs.	As the momentum increases, the OAM of the created particles also grows, which indicate that the spin entanglement is affected by the OAM.

\section{Conclusions and outlook}\label{summary}

In this work, we investigate vortex structures and entanglement properties in multiphoton pair production in circularly polarized fields through the two-level model. Our analysis reveals that spin configurations, topological charges, and phase singularity play crucial roles in the particle distributions and entanglement characteristics of the produced pair. The main conclusions are summarized as follows.

Spin configurations strongly influence both the particle distributions and the vortex toplogical charges. For the four-photon process at $\omega = 0.55m$, the spin configuration $S_z = 1$ has the highest probability and is associated with three topological charges, resulting in smallest ring radius in particle distributions. The spin configuration $S_z = -1$ corresponds to five topological charges and possesses the lowest probability, which leads to the largest ring radius in the particle distributions. Frequency increase causes a step-like decrease in topological charges. For multiphoton rings that dominate pair production, the variation of topological charges with field frequency can be understood through energy conservation and angular momentum conservation. In the low-momentum regime, a tunneling assisted mechanism becomes significant in the range of frequency $n\omega < 2m_*$ for  $n$-photon channel.	Pair entanglement is strongly dependent on momentum. As momentum increases, the number of photons absorbed and the OAM modes both increase, leading to a significant increase in particle entanglement.

In our investigation, we have extended the production of vortex states particles to the field of vacuum pair production. The method of measuring the spin correlation of entangled pair through secondary scattering of the particles on two independent targets is also expected to be applied to the study of entanglement in strong-field QED pair production processes~\cite{Gao:2025kdi}.

\begin{acknowledgments}
We are grateful to Q. Z. Lv, L. N. Hu, and Z. L. Li for helpful discussions.
This work is supported by the National Natural Science Foundation of China (NSFC) under Grants No. 12375240, No. 12535015, No. 12575261, and No. 12447179. L. J. Li acknowledges also support from the Youth Science and technology Foundation of Gansu Province  (Grant No. 24JRRA276).
\end{acknowledgments}

\end{document}